# Investigating the Acoustical Properties of Carbon Fiber-, Glass Fiber-, and Hemp Fiber-Reinforced Polyester Composites

Mohammad Mehdi Jalili,[1] Seyyed Yahya Mousavi,[2] Amir Soheil Pirayeshfar[1]

[1]Department of Polymer Engineering, Science and Research Branch, Islamic Azad University, Tehran, Iran

[2]Department of Polymer Engineering, Tehran South Branch, Islamic Azad University, Tehran, Iran

**Wood is one of the main materials used for making musical instruments due to its outstanding acoustical properties. Despite such unique properties, its inferior mechanical properties, moisture sensitivity, and time- and cost-consuming procedure for making instruments in comparison with other materials (e.g., composites) are always considered as its disadvantages in making musical instruments. In this study, the acoustic parameters of three different polyester composites separately reinforced by carbon fiber, glass fiber, and hemp fiber are investigated and are also compared with those obtained for three different types of wood specimens called poplar, walnut, and beech wood, which have been extensively used in making Iranian traditional musical instruments. The acoustical properties such as acoustic coefficient, sound quality factor, and acoustic conversion factor were examined using some non-destructive tests based on longitudinal and flexural free vibration and also forced vibration methods. Furthermore, the water absorption of these polymeric composites was compared with that of the wood samples. The results reveal that the glass fiber-reinforced composites could be used as a suitable alternative for some types of wood in musical applications while the carbon fiber-reinforced composites are high performance materials to be substituted with wood in making musical instruments showing exceptional acoustical properties. POLYM. COMPOS., 00:000–000, 2014. © 2014 Society of Plastics Engineers**

## INTRODUCTION

Acoustical characteristics of wood make it ideal for using it in a variety of musical instruments such as guitar and violin soundboard and for making clarinets, oboes, and drum sticks, just to mention a few. Spruce wood is usually used in top plates of violins and cellos because it has exceptional resonant quality and it is favored for soundboards whereas curly maple is used for their backs [1–4]. Among other species, poplar, walnut, and beech wood have been extensively used in making Iranian traditional musical instruments called Santour, Tar, and Barbat [5, 6]. Despite such unique acoustical properties of various species of wood, wood exhibits some drawbacks in such applications. For example, no desirable sound could be produced in humid conditions, and the existence of tie within the wood is inevitable. Furthermore, making a musical instrument from wood is a time- and cost-consuming procedure, and the properties are not exactly identical in all over the wood specimen [6, 7]. Some investigations showed that using polymeric composites instead of wood can overcome such problems [8, 9]. Among these composites, high performance fiber-reinforced composites not only can be tailored based on pre-determined properties but also offer high physical and mechanical properties during their service life [9–11]. Based on our literature review, just a few works have been performed examining acoustical properties of fiber-reinforced composites such as the works presented in Refs. [12–14]. Although some attempts have been made in this field, using of composites in musical application has not been yet well documented.

Acoustical properties of materials could be accurately estimated via non-destructive tests (NDTs) causing no damage and undesirable effects on the samples. One of the promising NDT methods is called resonant vibration method in which the test specimen is mechanically vibrated in a torsional, transverse, or longitudinal vibration mode over a range of frequencies [15–18]. This method has been used by some authors to investigate and analyze the acoustical performance of various species of wood [19–21].

In this work, we examine the acoustical properties of the prepared carbon fiber-, glass fiber-, and hemp fiber-

*Correspondence to*: Mohammad Mehdi Jalili;
e-mail: m.jalili@srbiau.ac.ir
Contract grant sponsor: Islamic Azad University-Science and Research Branch (IAUSRB).

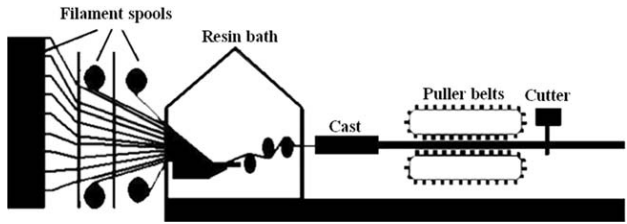

FIG. 1. The schematic presentation of pultrusion method used for the preparation of the fiber reinforced composites.

polyester composites using some resonant NDTs based on longitudinal and flexural free vibration and also forced vibration methods. We also compare acoustical properties of three different types of wood specimens called poplar, walnut, and beech wood, which have been extensively used in making Iranian traditional musical instruments, with those of the fiber-reinforced composites.

## EXPRIMENTAL

### Materials

In this article, isophthalic unsaturated polyester resin, Boshpol 751129, was purchased from Bushehr Chemical Industries (domestic company). Methyl ethyl ketone peroxide (1 wt%) (Pamookaleh, Iran) and 0.9 wt% cobalt octoate (Saveh Chemicals, Iran) were also added to the polyester resin according to the data sheet recommendations as initiator and accelerator, respectively. Carbon fiber (T300) from Troyca Company, Glass fiber (WR3) from Camelyaf Company (Turkey), and hemp fiber from domestic company were provided and used as fiber reinforcements, separately. Also, poplar wood, walnut wood, and beech wood were provided from northern forests of Iran. According to the material technical data sheets, density of carbon fiber, glass fiber, hemp fiber, and polyester resin were considered as 1.7, 2.4, 1.3, and 1.05 g/cm$^3$, respectively.

### Procedure of Sample Preparation

Pultrusion method was used to prepare the specimens (Schematic presentation of this method is given in Fig. 1). Through this method, rod-shaped fiber composite specimens with diameter of 9 mm were produced and then cut in the appropriate sizes for different analyses. In this study, fiber volume percent, determined by counting the number of filaments and measuring the weight of each single filament with a specific length [22, 23], was 70, 65, and 60 for carbon fiber-, glass fiber-, and hemp fiber-reinforced polyester composites, respectively.

A sample of pure polymer matrix without any fiber was also prepared and subjected to all NDT methods to measure its viscoelastic properties. But, it did not show distinguishable first modes of vibration similar those the fiber-reinforced polymer composites did. In fact, it could not vibrate solely in the absent of reinforcing fibers. Hence, we could not perform the NDT methods and obtain accurate results for the neat polymer matrix in this research.

## CHARACTERIZATION METHODS

### Non-destructive Longitudinal Free Vibration Test

To implement this technique, the set-up as shown in Fig. 2 was prepared. First, each test specimen was hold from its center and was hit by a wooden hammer at the end of specimen. To analyze the acoustic response of the specimen, a microphone was positioned in the other side of sample. Subsequently, the response vibrating sound was recorded by Audacity software as a wave-format file. A sound wave comprises of three components namely loudness, frequency, and time. To display each sound in terms of its components, all the recorded sounds were analyzed by the means of Fast Fourier Transform (FFT) using MATLAB V.7.1.

Meanwhile, wood specimens were cut as rectangular test specimens with dimensions of 2.5 × 2.5 × 30 cm$^3$. All measurements were performed on a set of five different replicates of composite or wood specimens.

### Mathematical Calculations Based on Longitudinal Free Vibration Test

Generally, ultrasonic velocity in a specimen could be determined from Eq. 1 [15, 24]:

$$V = f \times \lambda, \qquad (1)$$

where $V$ is ultrasonic velocity in a specimen, and $f$ is resonance frequency of the first mode of vibration when the vibration resonance occurs in the specimen for the first time. Also, $\lambda$ is wave length and can be calculated from Eq. 2 [15, 24]:

$$\lambda = \frac{2L}{n}, \qquad (2)$$

where $L$ is the length of specimen and $n$ is the number of resonance modes. Note that each vibrating material possesses infinitive modes of resonance. For the first mode of vibration, $n$ is equal to 1; therefore, the wave length of first mode of vibration can be obtained as follows:

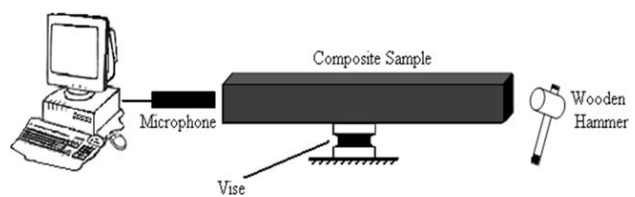

FIG. 2. The set-up of non-destructive longitudinal free vibration test.

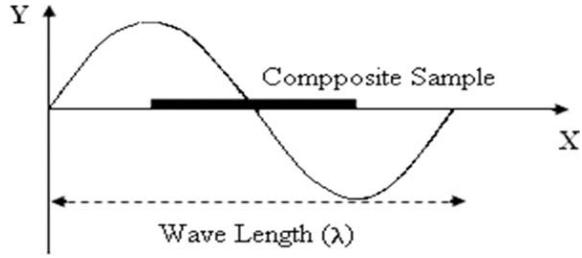

FIG. 3. The position of the node and anti-nodes, on the first vibrational mode, in a specimen hit by a wooden hammer at the end of specimen.

$$\lambda = 2L. \quad (3)$$

In fact, there is a node on the vibration wave at the center of the specimen combined with two anti-nodes at the both ends of the specimen as shown in Fig. 3. According to the positions of the node and the two anti-nodes corresponding to the first mode of vibration, the wave length is equal to the twice of the specimen length.

After calculating ultrasonic velocity ($V$), longitudinal elastic modulus as well as specific longitudinal elastic modulus ($S$) could be determined according to Eqs. 4 and 5, respectively [15, 24]:

$$E = \rho V^2, \quad (4)$$

where $E$ is elastic modulus, and $\rho$ is density.

$$S = \frac{E}{\rho}, \quad (5)$$

where $\rho$ is the specific density.

Acoustic coefficient of the vibrating body could be also calculated as expressed in Eq. 6 [24]:

$$K = \left(\frac{E}{\rho^3}\right)^{0.5}, \quad (6)$$

where $K$ is the acoustic coefficient of the vibrating body, $E$ is the longitudinal elastic modulus, and $\rho$ is the specific density.

After each impulse the sample starts vibrating. Due to the internal friction, the vibration energy reduces after a while. This attenuation occurs due to the material damping which is affected by the type of the materials. Therefore, decrement in vibration energy as a function of time could be a relevant key factor to determine the damping factor (tan $\delta$) of materials, as expressed in Eq. 7 [15, 24]:

$$\tan \delta = \frac{\lambda'}{\pi}, \quad (7)$$

where $\lambda$ is logarithmic vibrating decrement factor which could be calculated by Eq. 8 [15, 24]:

$$\lambda' = \left(\frac{1}{n}\right) \ln \left|\left(\frac{X_1}{X_{n+1}}\right)\right|, \quad (8)$$

where $n$ is time parameter, $X_1$ and $X_{n+1}$ are the first and ($n + 1$)th amplitude of vibration, respectively (see Fig. 4).

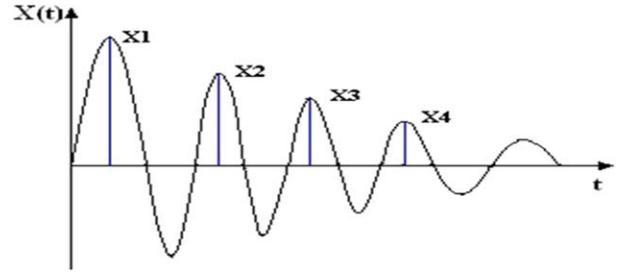

FIG. 4. The schematic view of amplitude decrement of the first mode of vibration through time. [Color figure can be viewed in the online issue, which is available at wileyonlinelibrary.com.]

Sound quality factor ($Q$) and acoustic conversion efficiency (ACE) which are both inversely proportional to tan $\delta$ are defined as follows [25]:

$$Q = \frac{1}{\tan \delta} \quad (9)$$

$$\text{ACE} = \frac{K}{\tan \delta}, \quad (10)$$

where $K$ is the acoustic coefficient of the vibrating body.

*Non-destructive Flexural Free Vibration Test*

To implement this technique, the set-up as shown in Fig. 5 was prepared. First, each test specimen was placed on two elastic jaws for the prevention of vibration damping. Then it was hit by a wooden hammer at the end of the specimen with a perpendicular impulse. A microphone was also positioned at the same position in the other side of the sample. To analyze the acoustic response of the specimen, the response vibrating sound was recorded by Audacity software as a wave-format file. Finally, all the recorded sounds were analyzed by the means of the FFT using MATLAB V.7.1.

*Mathematical Calculations Based on Flexural Free Vibration Test*

In flexural free vibration, the elastic modulus can be calculated according to the Timoshenko theory. This method was proposed in 1989 by Bordonne obtain the elastic and shear moduli of materials as a fast and confident approach [26]. According to the Timoshenko theory, the specific elastic modulus (the ratio of elastic modulus to the specific density) can be obtained by a linear regression on the values of $a_k$ and $b_k$ parameters as expressed in *Eq. 11*. The $a_k$ and $b_k$ parameters can be calculated from the Fast Fourier Transform for the frequency of $k$th vibration mode according to *Eqs. 12 and 13*:

$$a_k = \left(\frac{E}{\rho}\right) - \left[\frac{E}{(K \times G)}\right] b_k \quad (11)$$

$$a_k = \frac{[4\pi^2 L^2 f_k^2 (1 + \alpha F_{1k})]}{(\alpha X_k)} \quad (12)$$

$$b_k = \frac{[4\pi^2 L^2 f_k^2 F_{2k}]}{X_k}, \quad (13)$$

where $E$ is the elastic modulus, $\rho$ is the density, $K$ is a shape factor (in this research is equal to 0.833), $G$ is the shear modulus, $L$ is the specimen's length, and $f_k$ is the frequency of $k$th vibration mode obtained from analyzing the Fast Fourier Transform.

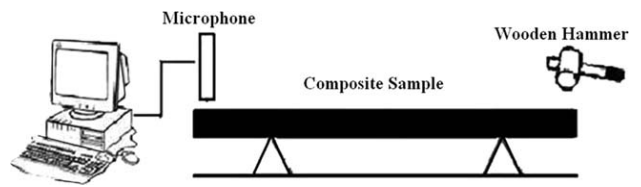

FIG. 5. The set-up of non-destructive flexural free vibration test.

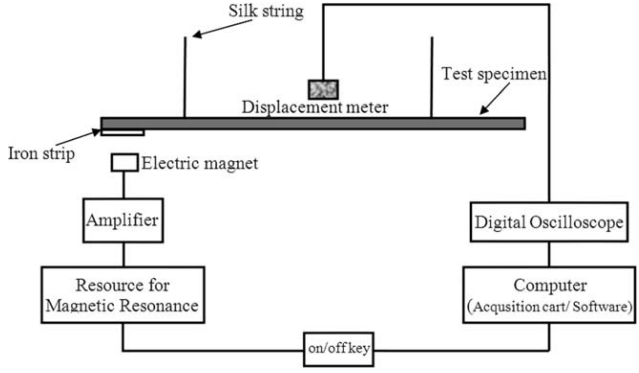

FIG. 6. The scheme of how flexural forced vibrational test works.

Furthermore, $\alpha$ is determined as follows:

$$\alpha = \frac{I}{AL^2}, \quad (14)$$

where $I$ is the moment of inertia, $A$ is the cross area, $L$ is the specimen length.

$X_k$, $F_{1k}$, and $F_{2k}$ in *Eqs. 12 and 13* can be also obtained from the below equations:

$$X_k = m_k^4 \quad (15)$$

$$F_{1k} = \Theta^2(m_k) + 6\Theta(m_k) \quad (16)$$

$$F_{2k} = \Theta^2(m_k) - 2\Theta(m_k) \quad (17)$$

$$m_k = \frac{(2k+1)\pi}{2} \quad (18)$$

$m_1 = 4.73$, $m_2 = 7.8532$, $m_3 = 10.9956$, …

$$\Theta(m_k) = \frac{[m_k \tan(m_k) \tanh(m_k)]}{[\tan(m_k) - \tanh(m_k)]}. \quad (19)$$

In the flexural free vibration test, damping factor (tan $\delta$), sound quality factor ($Q$), and acoustic conversion efficiency (ACE) can be calculated the same as those in the longitudinal free vibration test. However, three peaks should be separately analyzed in this method and therefore three separate values of damping factor are calculated. The average of the three obtained values was recorded and used as final damping factor.

*Non-destructive Forced Vibration Test*

The scheme of the forced vibration test is shown in Fig. 6. As it can be seen, a small iron strip was adhered to one of the ends of the sample. As the sample must have vibrated freely without any disturbance, as Fig. 6 shows, each test specimen nodes were placed in two silk string rings. It should be noted that the places of the string rings are very important and as we only analyze the first mode of vibration, they should be exactly on the nodes of the specimen's first mode of vibration. Otherwise, there would be some disturbance on the vibration of the sample. The places of the nodes of the first vibrational mode of the sample have been shown in Fig. 7. After the sample was placed appropriately, an electrical magnet was set exactly below the iron strip. As the on and off frequencies of the electrical magnet are completely changeable, it is possible to determine the resonance frequencies of the specimen just by changing the on and off frequencies of the electrical magnet. So, the on and off frequencies of the electrical magnet were increased from zero until the frequency that the microphone displacement meter, which is connected to the computer, shows the maximum vibrational amplitude. It should be noted that vibrational amplitude of the specimen would be maximum in the resonance modes of vibration. So, in this method, the first resonance frequency of the specimen would be determined by detecting the vibrational amplitude of the specimen. When the vibrational amplitude of the sample becomes the maximum, the frequency of the electrical magnet is the first resonance frequency of the specimen, which is obtained in the previous works by the FFT analysis of the recorded sounds. After detecting the first resonance frequency of the sample, the electrical magnet currency would be turned off suddenly. Therefore, the sample vibration would decrease until that it becomes zero. The damping factor of the specimen would be measured by analyzing the intensity of this decrement [20]. To analyze the stated inputs and also to calculate the results, Matlab7.1 and Microsoft Excel 2003 were used [20].

*Mathematical Calculations Based on Forced Vibration Test*

According to the ASTM 1548-02, the dynamic elastic modulus of isotropic materials (this is also acceptable for orthotropic materials like composite samples which have been built in this research) can be calculated from the first mode of vibration which is obtained from the forced vibration test according to the below equation [20, 21]:

$$E_{da} = \left[\frac{mf_f^2}{b}\right]\left[\frac{L^3}{h^3}\right]T_1, \qquad (20)$$

where $E_{da}$ is the dynamic elastic modulus (Pa), $f_f$ is the first resonance frequency obtained from the forced vibration test (Hz), $m$ is the sample weight (kg). $b$, $h$, and $L$ are specimen's length, width, and height, respectively. $T_1$ is the corrective factor which must be entered if the length to height ratio of the specimen was more than 20 (like the samples prepared for this research). The $T_1$ factor is calculated according to Eq. 21:

$$T_1 = \left\{1.000 + 6.585\left(\frac{h}{L}\right)^2\right\}. \qquad (21)$$

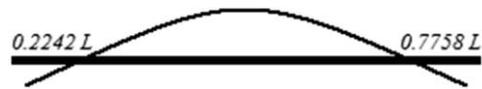

FIG. 7. The place of nodes on the first vibrational mode.

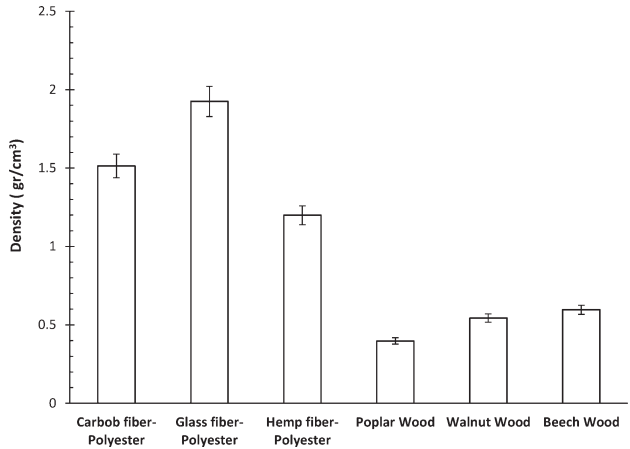

FIG. 8. The measured density for the carbon fiber-, glass fiber-, and hemp fiber-reinforced polyester composites as well as the poplar, walnut, and beech wood specimens.

*Water Absorption Test*

The samples were cut in appropriate dimensions and then they were heated for 40 min in a drying oven with 80°C temperature to be dried completely. The specimens were weighed right after coming out of the oven using a Mettler Model College 150 balance with precision of 0.0001 g. Then, each sample is immersed into distilled water at the ambient temperature for a long time and in certain periods of time is weighed again after being dried with a soft cloth. Finally, the water absorption of each specimen is calculated from the equation below:

$$M = \left[\frac{(M_t - M_0)}{M_0}\right] \times 100, \qquad (22)$$

where $M_0$ is the primary weight before immersion, $M_t$ is the weight of the sample after the immersion at time $t$, and $M$ is the percentage of water absorption in different periods of immersion.

## RESULTS AND DISCUSSION

As previously mentioned, the density of a material plays a significant role in its acoustical properties. In general, light materials can retain the vibration energy more than materials having higher density can do. According to *Eq. 6*, when the density is low, like the density of wood, the high acoustic coefficient is reasonably expected. However, the density of fiber reinforced composites is much higher than that of wood due to their high density fibers. The measured density for carbon fiber-, glass fiber-, and hemp fiber-reinforced polyester composites as well as for the examined wood specimens (i.e., walnut, poplar, and beech wood) is shown in Fig. 8. As it can be seen, the density of fiber composites is much greater than that of wood samples particularly when the glass fiber is used. According to *Eq. 6*, materials with such higher density than wood would be expected to represent poorer acoustical properties, unless they possess greater elastic modulus than that of wood. The longitudinal and flexural free vibration and also forced vibration NDTs were therefore conducted to determine the elastic modulus of the composites and wood specimens.

*Longitudinal Properties of Specimens Obtained by Longitudinal Free Vibration Test*

By obtaining the frequency values of the first mode of vibration and using *Eq. 1*, the sound velocity of each sample can be calculated. Table 1 shows the calculated values of the sound velocity for the fiber reinforced composites together with wood specimens. It seems that the composites reinforced either with glass fiber or hemp fiber represent a sound velocity value almost the same as that of wood specimens. Among these, the carbon fiber reinforced composite distinguishingly reveals the highest value so that the speed of sound in this composite is found to be surprisingly more than two times greater than that in the wood specimens.

The elastic modulus values of all test specimens calculated based on the obtained longitudinal ultrasonic velocities are also given in Table 1. Although Table 1 shows much greater values of modulus for all reinforced composites than those measured for the three different kinds of wood, the specific modulus values (when density is considered) of the glass fiber- and hemp fiber-reinforced composites are close to the specific modulus of wood. However, the specific modulus of the carbon fiber-reinforced composites is still much higher than that of wood specimens.

Acoustic coefficient values, calculated from *Eq. 6*, are also given in Table 1. Generally, the wood specimens show higher values than the polymeric ones. Among the polymeric composites, only the carbon fiber-polyester composite possesses an acoustic coefficient value (238.56) approximately close to that of walnut and beech wood specimens.

TABLE 1. Acoustic properties of the carbon fiber-, glass fiber-, and hemp fiber-reinforced polyester composites as well as the poplar, walnut, and beech wood specimens obtained from the non-destructive longitudinal free vibration test.

|  | Carbon fiber-reinforced polyester composite | Glass fiber-reinforced polyester composite | Hemp fiber-reinforced polyester composite | Poplar wood | Walnut wood | Beech wood |
|---|---|---|---|---|---|---|
| Longitudinal ultrasonic velocity (m/s) | 11,423 | 4,726 | 4,156 | 4,763 | 4,037 | 4,974 |
| Standard deviation | 9 | 8 | 23 | 27 | 26 | 19 |
| Longitudinal elastic modulus (GPa) | 197.5 | 43.0 | 17.4 | 9.0 | 8.8 | 13.0 |
| Standard deviation | 0.3 | 0.2 | 0.2 | 0.1 | 0.1 | 0.1 |
| Longitudinal specific elastic modulus (GPa) | 130.5 | 22.3 | 17.3 | 22.7 | 16.3 | 21.8 |
| Standard deviation | 0.2 | 0.1 | 0.2 | 0.3 | 0.2 | 0.2 |
| Acoustic coefficient | 238.6 | 77.6 | 130.3 | 378.5 | 235.1 | 247.6 |
| Standard deviation | 0.2 | 0.1 | 0.7 | 2.2 | 1.5 | 1.0 |
| Damping factor (tan $\delta$) | 0.0009 | 0.0026 | 0.0170 | 0.0069 | 0.0083 | 0.0108 |
| Standard deviation | 0.0003 | 0.0002 | 0.0002 | 0.0003 | 0.0003 | 0.0003 |
| Quality factor | 1111.1 | 384.6 | 58.8 | 144.9 | 120.5 | 92.3 |
| Standard deviation | 198.4 | 34.0 | 0.7 | 6.0 | 4.2 | 2.5 |

Basically, the released sound from a wood specimen disappears after a while due to the damping of vibration energy within the specimen. Therefore, if a test specimen possesses a minimum damping factor value (approx. zero), it can vibrate for a long time after disconnecting the vibration source [15, 27]. In other words, damping factor is an indicative of the internal friction coefficient which is desired to be as less as possible for a material which would be used in sound boxes and sound boards of acoustic musical tools. Based on this factor, the ranking of the composites from high to low would be as follows: hemp-reinforced composites, wood composites (i.e., poplar, walnut, and beech wood which are naturally made of lignin resin and cellulose fibers), glass fiber-reinforced composites, carbon fiber-reinforced composites. It means that although polymeric composites are not perfectly elastic and therefore the vibration energy is expected to reduce after a while, the composites reinforced with carbon fiber and glass fiber can vibrate for a longer time after disconnecting the vibration supply whereas the wood specimens cannot. *Equation 9* shows that damping factor (tan $\delta$) is related to the quality factor $Q$ (inverse of the quality factor $Q$) which is a descriptor of the sound quality of a material. This descriptor shows that samples with high damping factor are associated with a poor acoustic quality. It has been proved that the more the quality factor is, the less wave damping the wood has. On count of the fact and by comparing the damping factor acquired for the wood specimen, it can be concluded that the sound quality factor of the composites reinforced by carbon fiber or glass fiber is much higher than $Q$ factor of different species of wood commonly used in making musical instruments. As an illustration, $Q$ factor of other types of wood such as maple and spruce which are commonly used in making guitar and violin varies approximately from 80 to 135 [5].

The higher values obtained for the $Q$ factor of such composites work as one of the influential factors making them as prospective choices in making musical instruments.

*The Flexural Properties of Specimens Obtained by Flexural Free Vibration Test*

In this approach, the first three vibration peaks were analyzed as described earlier, and eventually the obtained values of sound velocity were used to calculate the values of flexural elastic modulus and specific flexural elastic modulus for the fiber reinforced composites together with the wood specimens. These data are given in Table 2. According to the results, there is no significant difference between the specific flexural modulus values (considering density) of the glass fiber- and hemp fiber-reinforced composites and those obtained for wood specimens, which is similar to what observed in longitudinal free vibration method. However, the obtained value for the carbon fiber-reinforced composites obviously distinguishes such composites from other test specimens suggesting that much higher acoustical properties in comparison with wood specimens can be expected.

Meanwhile, Table 2 shows acoustic coefficient values for all the composites and the wood specimens. The trend is still the same as what previously observed in longitudinal free vibration method showing that the wood specimens possess higher values than the polymeric ones, and only the carbon fiber-polyester composite can be considered as an alternative with an acceptable acoustic coefficient.

TABLE 2. Acoustic properties of the carbon fiber-, glass fiber-, and hemp fiber-reinforced polyester composites as well as the poplar, walnut, and beech wood specimens obtained from the non-destructive flexural free vibration test.

| | Carbon fiber-reinforced polyester composite | Glass fiber-reinforced polyester composite | Hemp fiber-reinforced polyester composite | Poplar wood | Walnut wood | Beech wood |
|---|---|---|---|---|---|---|
| Flexural elastic modulus (GPa) | 189.1 | 38.6 | 12.6 | 9.8 | 9.8 | 14.0 |
| Standard deviation | 0.2 | 0.1 | 0.3 | 0.1 | 0.1 | 0.2 |
| Flexural specific elastic modulus (GPa) | 124.9 | 20.1 | 12.5 | 24.6 | 18.1 | 23.5 |
| Standard deviation | 0.2 | 0.3 | 0.3 | 0.3 | 0.2 | 0.3 |
| Acoustic coefficient | 233.4 | 73.6 | 110.8 | 393.9 | 247.9 | 256.7 |
| Standard deviation | 0.2 | 0.3 | 0.8 | 2.3 | 1.3 | 1.6 |
| Damping factor (tan $\delta$) | 0.0017 | 0.0035 | 0.0210 | 0.0103 | 0.0118 | 0.0151 |
| Standard deviation | 0.0003 | 0.0002 | 0.0002 | 0.0002 | 0.0003 | 0.0007 |
| Quality factor | 588.2 | 285.7 | 47.6 | 97.1 | 84.74 | 66.2 |
| Standard deviation | 113.8 | 18.9 | 0.5 | 1.7 | 2.2 | 3.3 |

The obtained values of damping factor and sound quality factor ($Q$ factor) are also demonstrated in Table 2. As seen, damping factor of carbon fiber- and glass fiber-reinforced composites is much less than that of wood specimens resulting in much higher $Q$ factor values for carbon fiber- and glass fiber-reinforced composites compared with other test specimens.

## The Flexural Properties of Specimens Obtained by Forced Vibration Test

The flexural properties of all specimens were also examined by another NDT based on forced vibration analysis to rest assured that they are measured correctly. The ultimate acoustic results for all composite and wood specimens calculated by forced vibration NDT have been shown in Table 3.

As it can be seen, a good agreement can be seen between the results measured with forced vibration NDT and those calculated with flexural free vibration NDT. To explain the causes of the small differences, it should be noted that different methods cannot exactly result in identical values due to their basic theoretical assumptions and/or practical conditions. Also, after that the specimens went through the flexural free vibration NDT, they were prepared for the forced vibration NDT. To prepare the samples for the forced vibration NDT, some cutting and abrasion processes were involved. Although this preparation method might introduce some stresses in the specimens and therefore decrease their properties to some extent, it is the best method to size such high-modulus materials to our best knowledge.

Figure 9 illustrates acoustic conversion efficiency (ACE) of the carbon fiber-, glass fiber-, and hemp fiber-reinforced polyester composites as well as the poplar, walnut, and beech wood specimens obtained from longitudinal and flexural free vibration and also forced vibration NDTs. It should be noted that acoustic conversion efficiency (ACE) could be considered as a key parameter to evaluate acoustic performance of a material as it comprises of both significant acoustic factors (i.e., acoustic coefficient, $K$, and sound quality factor, $Q$). Roohnia et al. used ACE values to rank acoustic performance of wide range of wood spices and expressed an excellent agreement between calculated ACE values and results obtained from practical experiences [27]. As it can be seen from Fig. 9, it is obvious that the hemp fiber-reinforced composites could not be a suitable alternative for wood in making musical instrument due to its low acoustic conversion efficiency. Reinforcing the composites by glass fiber, however, could provide a synthetic material for replacing with some types of wood in musical applications. Another finding of ACE results is that the carbon fiber-reinforced composites are high performance materials to be substituted with wood in making music instruments showing exceptional acoustical properties.

TABLE 3. Acoustic properties of the carbon fiber-, glass fiber-, and hemp fiber-reinforced polyester composites as well as the poplar, walnut, and beech wood specimens obtained from the non-destructive forced vibration test.

| | Carbon fiber-reinforced polyester composite | Glass fiber-reinforced polyester composite | Hemp fiber-reinforced polyester composite | Poplar wood | Walnut wood | Beech wood |
|---|---|---|---|---|---|---|
| Flexural elastic modulus (GPa) | 180.7 | 35.8 | 11.2 | 9.1 | 9.5 | 12.9 |
| Standard deviation | 1.0 | 0.3 | 0.2 | 0.2 | 0.1 | 1.5 |
| Flexural specific elastic modulus (GPa) | 119.3 | 18.6 | 11.1 | 22.9 | 17.5 | 21.6 |
| Standard deviation | 0.6 | 0.1 | 0.2 | 0.4 | 0.2 | 2.5 |
| Acoustic coefficient | 228.2 | 70.8 | 104.4 | 379.9 | 244.0 | 246.3 |
| Standard deviation | 0.6 | 0.3 | 1.1 | 3.5 | 1.3 | 15.5 |
| Damping factor (tan $\delta$) | 0.0017 | 0.0041 | 0.0260 | 0.0110 | 0.0113 | 0.0167 |
| Standard deviation | 0.0001 | 0.0002 | 0.0002 | 0.0003 | 0.0001 | 0.0003 |
| Quality factor | 588.2 | 243.9 | 38.5 | 91.3 | 88.3 | 60.0 |
| Standard deviation | 43.8 | 11.2 | 0.4 | 2.3 | 0.2 | 1.2 |

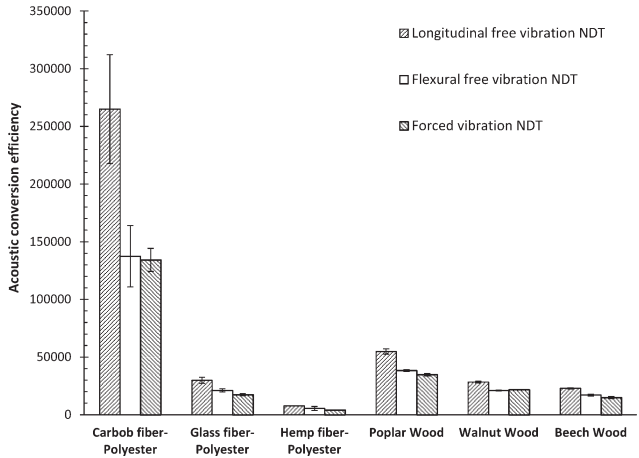

FIG. 9. Acoustic conversion efficiency of the carbon fiber-, glass fiber-, and hemp fiber-reinforced polyester composites as well as the poplar, walnut, and beech wood specimens obtained from various non-destructive resonant tests.

*Water Absorption*

The high water absorption of wood is one of the biggest problems of wooden musical instruments. As wood absorbs the moisture in humid environments, its acoustic and vibrational properties diminish, and this jeopardizes the sound quality of the musical instruments. Furthermore, when the wet wood is drying, some changes in the shape of the musical instrument might occur and it could be mostly observed as some little bending in the neck of stringed musical instruments.

It has been shown that mechanical properties of composite materials would decrease after being either exposed to a humid environment or immersed in water for a while and as a result their acoustical properties would be negatively affected. It has been also revealed that water immersion condition leads to a greater decrease in acoustical properties of composite materials than humid environment does [28–30]. Therefore, in this research the harder condition was chosen and the samples were immersed in distilled water for 40 days. The water uptake of all specimens consisting of the carbon fiber-, glass fiber-, and hemp fiber-reinforced polyester composites as well as the poplar, walnut, and beech wood specimens was measured through time. The results are depicted in Fig. 10.

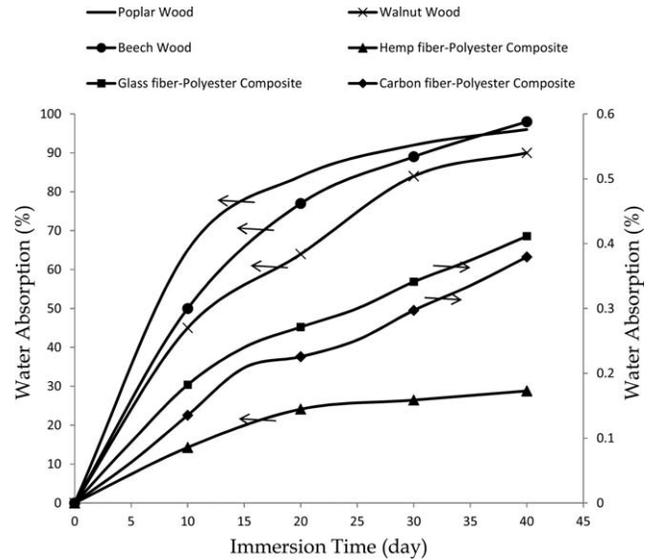

FIG. 10. Water absorption percentage of the carbon fiber-, glass fiber-, and hemp fiber-reinforced polyester composites as well as the poplar, walnut, and beech wood specimens after 40 days immersion (the arrow shows the vertical axis for each curve).

As it clearly can be seen, the water absorption of polymeric composites is greatly less than that of the wood samples. The water absorption of carbon fiber and glass fiber composites is negligible in comparison with those of the wood samples. Also the water absorption of hemp fiber reinforced polyester composite, although is greater than carbon fiber and glass fiber composites, is much less than that of the wood samples. Given that the water immersion test is a harder method than humid exposure test, the composite samples, especially carbon and glass composites, simply pass this test. By analyzing the water absorption percentages of polymeric composites and comparing them to those of the wood specimens, it can be easily concluded that the environmental humidity will have no (or negligible) effects on the musical instruments which have been made by polymeric composites specially carbon fiber- and glass fiber- polyester ones.

## CONCLUSIONS

This work compares acoustic parameters of three different polyester composites separately reinforced by carbon fiber, glass fiber, and hemp fiber with those of wood specimens. Poplar, walnut, and beech wood, which are mainly used in making Iranian traditional musical instruments, were selected as control wood specimens. In this article, the acoustical properties such as acoustic coefficient, sound quality factor, and acoustic conversion efficiency have been investigated using some NDTs based on longitudinal and flexural free vibration and also forced vibration methods. It is found that the results obtained from longitudinal free vibration method and those measured by flexural tests are in a good agreement for all the examined composites. The glass fiber-reinforced composites show an acoustic performance similar to what walnut wood does. The results also reveal that the carbon fiber-reinforced composite could be used as an improved tailored high performance alternative to be substituted with wood in making musical instruments showing exceptional acoustical properties.

Also, the water immersion results showed that the effect of environmental humidity on the sound quality of musical instruments which have been manufactured by the polymeric composites, especially carbon fiber- and glass fiber composites is negligible.